# Aggregating the response in time series regression models, applied to weather-related cardiovascular mortality

Pierre Masselot[1*], Fateh Chebana[1], Diane Bélanger[1,2], André St-Hilaire[1],

Belkacem Abdous[3], Pierre Gosselin[1,2,4], Taha B.M.J. Ouarda[1]

[1]*Institut National de la Recherche Scientifique, Centre Eau-Terre-Environnement, Québec, Canada;*

[2]*Centre Hospitalier Universitaire de Québec, Centre de Recherche, Québec, Canada;*

[3]*Université Laval, Département de médecine sociale et préventive, Québec, Canada;*

[4]*Institut national de santé publique du Québec (INSPQ), Québec, Canada.*

*Corresponding Author: masselot.pierre@gmail.com*

# Abstract

In environmental epidemiology studies, health response data (*e.g.* hospitalization or mortality) are often noisy because of hospital organization and other social factors. The noise in the data can hide the true signal related to the exposure. The signal can be unveiled by performing a temporal aggregation on health data and then using it as the response in regression analysis. From aggregated series, a general methodology is introduced to account for the particularities of an aggregated response in a regression setting. This methodology can be used with usually applied regression models in weather-related health studies, such as generalized additive models (GAM) and distributed lag nonlinear models (DLNM). In particular, the residuals are modelled using an autoregressive-moving average (ARMA) model to account for the temporal dependence. The proposed methodology is illustrated by modelling the influence of temperature on cardiovascular mortality in Canada. A comparison with classical DLNMs is provided and several aggregation methods are compared. Results show that there is an increase in the fit quality when the response is aggregated, and that the estimated relationship focuses more on the outcome over several days than the classical DLNM. More precisely, among various investigated aggregation schemes, it was found that an aggregation with an asymmetric Epanechnikov kernel is more suited for studying the temperature-mortality relationship.

## 1. Introduction

In environmental epidemiology, studies on the health effect of various environmental exposures, often rely on regression models applied to time series data (Gasparrini and Armstrong, 2013). Environmental exposure variables include atmospheric pollutant levels and temperature, while health issues include various diseases (Barreca and Shimshack, 2012; Blangiardo et al., 2011; Braga et al., 2002; Knowlton et al., 2009; Martins et al., 2006; Nitschke et al., 2011; Szpiro et al., 2014; Yang et al., 2015). In this context, the exposure-response relationship is complex since, among other reasons, the effect of exposure on health issues lasts several days. This is why models have often used exposure windows under the form of moving averages (MA, *e.g.* Armstrong, 2006) or distributed lags (DL, Schwartz, 2000). In particular, the latter has been extended to deal with nonlinear relationships (distributed lags nonlinear models, DLNM, Gasparrini et al., 2010) which is now widely used in weather-related health population studies (*e.g.* Phung et al., 2016; Vanos et al., 2015; Wu et al., 2013).

The health response, however, is almost always used directly as a daily time series. This could lead to several drawbacks in the regression models of environmental exposure on a health issue. First, the response to an exposure can also be spread across several days (Lipfert, 1993), which means that it would seem more realistic to consider a health time window in response to an associated exposure window. Second, health time series data used in epidemiologic studies are often noisy. The noise can conceal the true signal of the response to an exposure, especially in areas with small populations where the number of cases (mortality or morbidity) is low. Sources of noise include diverse organizational factors such as weekends and holidays (Suissa et al., 2014; Wong et al., 2009), slight changes in the definition of diseases (*e.g.* Antman et al., 2000) as well as behavioral and technological changes. In the end, the noise in the response can reduce the accuracy of the model and the conclusions (*e.g.* Todeschini et al., 2004).

In order to assess a more realistic relationship between an exposure and a health issue as well as reduce the noise impact in the health response, it is proposed to consider an aggregation window over time in the health response also, in addition to the exposure. More precisely, moving aggregation is considered here, *i.e.* the time step of data points in the obtained series remains the same, in opposition to aggregation where the time step of data points is reduced (*e.g.* from daily values to monthly values).

Aggregating the response series is expected to have two advantages: (1) better representing the spread of the health response to an exposure and (2) reducing the noise in the health series. Indeed, aggregated series are less sensitive to random perturbation in the data. An aggregated response should make regression models more robust to variations induced by noise, leading to more reliable relationship estimates. This idea is consistent with the results of Cristobal et al. (1987) in a non-time series context, which showed that pre-smoothing a response variable to remove noise leads to consistent estimates with low variance in linear regression. In a similar study, Sarmento et al. (2011) concluded that regression models are more robust to noise when both the response and the exposure are aggregated.

There have been few preceding cases of aggregated responses (Roberts, 2005; Sarmento et al., 2011; Schwartz, 2000b), but the regression models applied did not account for the specificities of an aggregated response. These specificities include the presence of extra autocorrelation in the residuals and a modification of their distribution. Therefore, the objective of the present paper is to introduce a general methodology dealing with an aggregated response. The methodology allows the use of a DLNM with an aggregated response and deals with the autocorrelation created by the aggregation. The exposure-response surface of a DLNM with aggregated response is then compared to the surface of a classical DLNM in order to assess the impact on the estimated relationship. In past studies, only the moving average (Roberts, 2005; Sarmento et al., 2011) and Loess (Schwartz, 2000b) have been considered to aggregate the response. In the present paper, other aggregations are considered, in particular Nadaraya-Watson kernel smoothing (Nadaraya, 1964; Watson, 1964) with different kernels including the Epanechnikov kernel (Epanechnikov, 1969) and an asymmetric kernel proposed in Michels (1992).

The paper is organized as follows. Section 2 introduces the proposed methodology for an aggregated response. Section 3 illustrates the methodology and its benefits by applying it on a weather-related cardiovascular mortality case. The methodology is first compared to models with a non-aggregated response and then, different aggregation strategies are compared. The results are discussed in section 4 and the conclusions are presented in section 5.

## 2. Methods

This section introduces the statistical methodology consisting in 1) performing a temporal aggregation on the response time series $y_t$; and 2) modelling the aggregated response $\tilde{y}_t$ according to an exposure $x_t$ through a regression model.

## 2.1. Aggregation of the response

The temporal aggregations considered in the proposed methodology are all local, *i.e.* the aggregation $\tilde{y}_t$ of a time series $y_t$ depends only on a subset of observations close to the current observation. In particular, linear local aggregations can be expressed as:

$$\tilde{y}_t = \sum_{i \in I} w_i y_{t+i} \tag{1}$$

where the $w_i$ are the weights attributed to each observation and $I$ is the aggregation window. The most common aggregation is the $H$-day centered moving average (MA) where $w_i = 1/H$, $I = \left\{-\frac{H-1}{2}; \ldots; \frac{(H-1)}{2}\right\}$ and $H$ (odd number) is the size of the window $I$. The main issue with MA is that the attributed weights are constant, giving equal importance to all the observations covered by the window $I$, even the farthest ones. This can result in small distortions in $\tilde{y}_t$ (Schwartz et al., 1996). The most common alternative aggregation is the Nadaraya-Watson kernel smoothing (Nadaraya, 1964; Watson, 1964) which generalizes the MA with the $w_i$ following a deterministic function $K(.)$ called "kernel". This allows for the weights to be non-constant. Although a number of kernels exist, the most popular is the Epanechnikov kernel (Epanechnikov, 1969) designed to minimize the squared error of the fit for a fixed window $I$. In addition, the asymmetric kernel of Michels (1992) is considered (thereafter called the Michels kernel). Unlike the MA and the Epanechnikov kernels, the Michels kernel is asymmetric (*i.e.* $w_i \neq w_{-i}$), giving more weights to future values ($i > 0$) in order to better represent physiological adaptation.

Classical aggregation includes past values ($i < 0$) to compute $\tilde{y}_t$. However, in the context of the response of a regression model, it seems more logical to explain only the future health response at time $t$. Therefore, all three aggregations discussed above (MA, Epanechnikov and Michels kernels) are considered using only future values (*i.e.* $I = \{0; ...; H-1\}$). The MA for future values attributes constant weights, the Epanechnikov kernel attributes decreasing weights from $i = 0$ to $i = H - 1$ and the Michels kernel attributes the maximum weight few days after the current one. Their shape is shown in Figure 1.

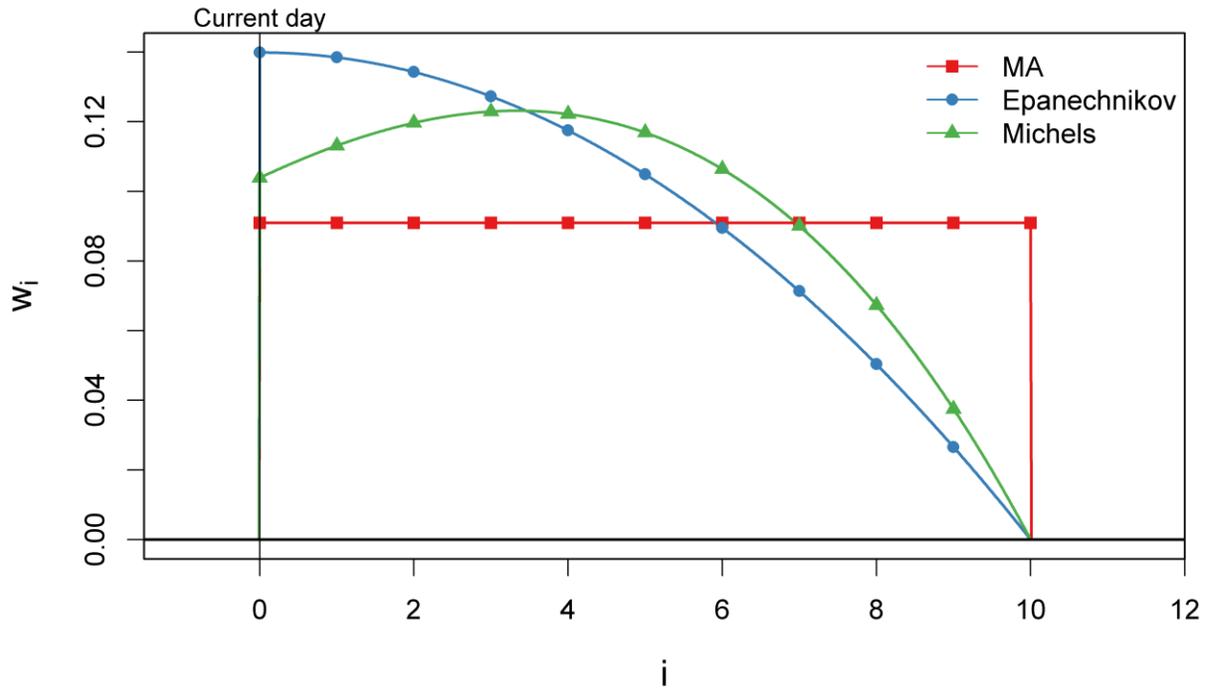

**Figure 1: Illustration of the weights $w_i$ versus $i$ in Eq. (1) where the weights are defined following the three types of kernels considered in the present paper: moving average (MA), Epanechnikov and Michels kernels. In this figure, the window size is set to $H = 11$.**

## 2.2. Regression model with aggregated response

The second step of the methodology is to use the aggregated response $\tilde{y}_t$ as the response of the general regression model:

$$\tilde{y}_t = \beta x_t + \epsilon_t \tag{2}$$

where $\beta$ is the regression coefficient and $\epsilon_t$ is the residual of the regression. The aggregated response $\tilde{y}_t$ raises two questionings for the model in Eq. (2): the probability distribution of residuals and the temporal dependence created by the aggregation.

When using a linear aggregation such as in Eq. (1), which is the case here, the distribution of $\epsilon_t$ can generally be considered Gaussian. This is justified by the central-limit theorem, which still holds for correlated variables (*e.g.* Billingsley, 1995, section 27), since the residuals are centered.

When temporal dependence is present in the response variable, it can be modelled through the residuals $\epsilon_t$. It is important to take it into account, since this violates the independence assumption of residuals, which increases the variance of estimators (*e.g.* Mizon, 1995). Regression models with serially correlated residuals are commonly known as time series regression (*e.g.* Choudhury et al., 1999). In the time series regression, the temporal dependence of residuals $\epsilon_t$ is modelled through an autoregressive-moving average model (ARMA, Box and Jenkins, 1976), *i.e.*

$$\epsilon_t = e_t + \sum_{k=1}^{p} \phi_k \epsilon_{t-k} + \sum_{l=1}^{q} \theta_l e_{t-l} \tag{3}$$

where $e_t$ does not have temporal dependency anymore, $p$ and $q$ are the orders of the ARMA model while the $\phi_k$ $(k = 1, ..., p)$ and $\theta_l$ $(l = 1, ..., q)$ are the model parameters. To estimate the ARMA model of Eq. (3), the regression model (Eq. 2) is first fitted in order to obtain a residual series on which the orders $p$ and $q$ are estimated. They are hereby estimated through a stepwise algorithm seeking to minimize the Akaike information criterion (AIC, Akaike, 1974; Hyndman and Khandakar, 2007). Minimizing the AIC allows a trade-off between goodness of fit and model complexity (high $p$ and $q$). However, other methods exist to estimate $p$ and $q$, such as simply analyzing the autocorrelation and partial autocorrelation functions or the extended autocorrelation function (EACF, Tsay and Tiao, 1984). The main advantage of the approach of Hyndman and Khandakar (2007) is that it is completely automated, which reduces the subjectivity of the process. Note that similarly to the AIC, the Bayesian information criterion (BIC, Schwarz, 1978) can also be used in the stepwise algorithm (we refer to Brewer et al., 2016 as well as Burnham and Anderson, 2004 for comparisons between AIC and BIC).

Finally, replacing the residual term in Eq. (2) by Eq. (3) leads to the full model used here, *i.e.*

$$\tilde{y}_t = \beta x_t + e_t + \sum_{k=1}^{p} \phi_k \epsilon_{t-k} + \sum_{l=1}^{q} \theta_l e_{t-l} \qquad (4)$$

The model in Eq. (4) is estimated with the maximum likelihood method (Pagan and Nicholls, 1976; Pesaran, 1973). This estimation method is more consistent than the generalized least squares approach (Aitken, 1935) and prewhitening (Cochrane and Orcutt, 1949), as explained by *e.g.* Mizon (1995) and Choudhury et al. (1999).

Note that the full model (Eq. 4) is linear. However, as stated in the introduction, it is also important to use lags of the exposure $x_t$ as well as allow the relationship to be nonlinear. The two aspects were combined in the DLNM approach and have proven useful (*e.g.* Gasparrini et al., 2015). Luckily, the surface of the DLNM is estimated by basis expansion which reduces the estimation to a linear model where the covariates are not the direct exposure anymore, but B-spline bases instead (mathematical details can be found in Gasparrini et al., 2010). Hence, the DLNM can be used as in Eq. (444) and the maximum likelihood estimator can be used as normal.

### 2.3. Performance assessment

Performance assessment is herein done through the $R^2$ criterion and cross-validation (CV, Stone, 1974). In order to fairly assess the performances, the criteria must be computed for the whole methodology (*i.e.* the two steps) rather than only for the regression step. Hence, the reference vector for these criteria is the response before aggregation $y_t$ (instead of the aggregated response $\tilde{y}_t$). This also allows for a fair comparison with classical models without an aggregated response. In addition, since the core of the methodology is to preprocess the response variable, many classical model comparison methods cannot be applied in the present case. For instance, Fisher tests are limited to nested models, and information criteria (AIC, BIC) cannot be directly applied because the responses used to compute the likelihood are different. The issue of different responses also forces us to compute the classical version of the $R^2$ based on the sum of squares of errors, instead of its generalized one based on the likelihood function (Magee, 1990).

Since the data used here are time series, the classical CV does not hold because it assumes independence between training and validation sets. Hence, it is recommended to use hv-block CV (Racine, 2000) which has been shown to perform well for time series data (Bergmeir and Benítez, 2012). The necessity to use the hv-block CV also justifies the use of local aggregations such as those introduced above, since the aggregation can be done independently in the training and validation sets of the CV.

## 3. Application and comparison

In Canada, cardiovascular diseases remain the main cause of mortality and put an increasing burden on the public health system (Wielgosz et al., 2009). It has already been shown that temperature affects cardiovascular mortality and morbidity (*e.g.* Bayentin et al., 2010; Bustinza et al., 2013; Masselot et al., 2018). Therefore, in order to efficiently organize private and public health service and mitigate the effect of temperature on cardiovascular diseases, it is important to understand every aspect of the relationship. To contribute to this goal, the proposed methodology, in which the response is aggregated, is applied to the common issue of temperature-related cardiovascular mortality in the census metropolitan area (CMA) of Montreal, Canada. In addition, a classical DLNM is applied to the same initial data in order to assess the importance of aggregating. Once the relevance of the proposed method is established, given the availability of a number of aggregation techniques, the issue of which one to consider is addressed by comparing several of them.

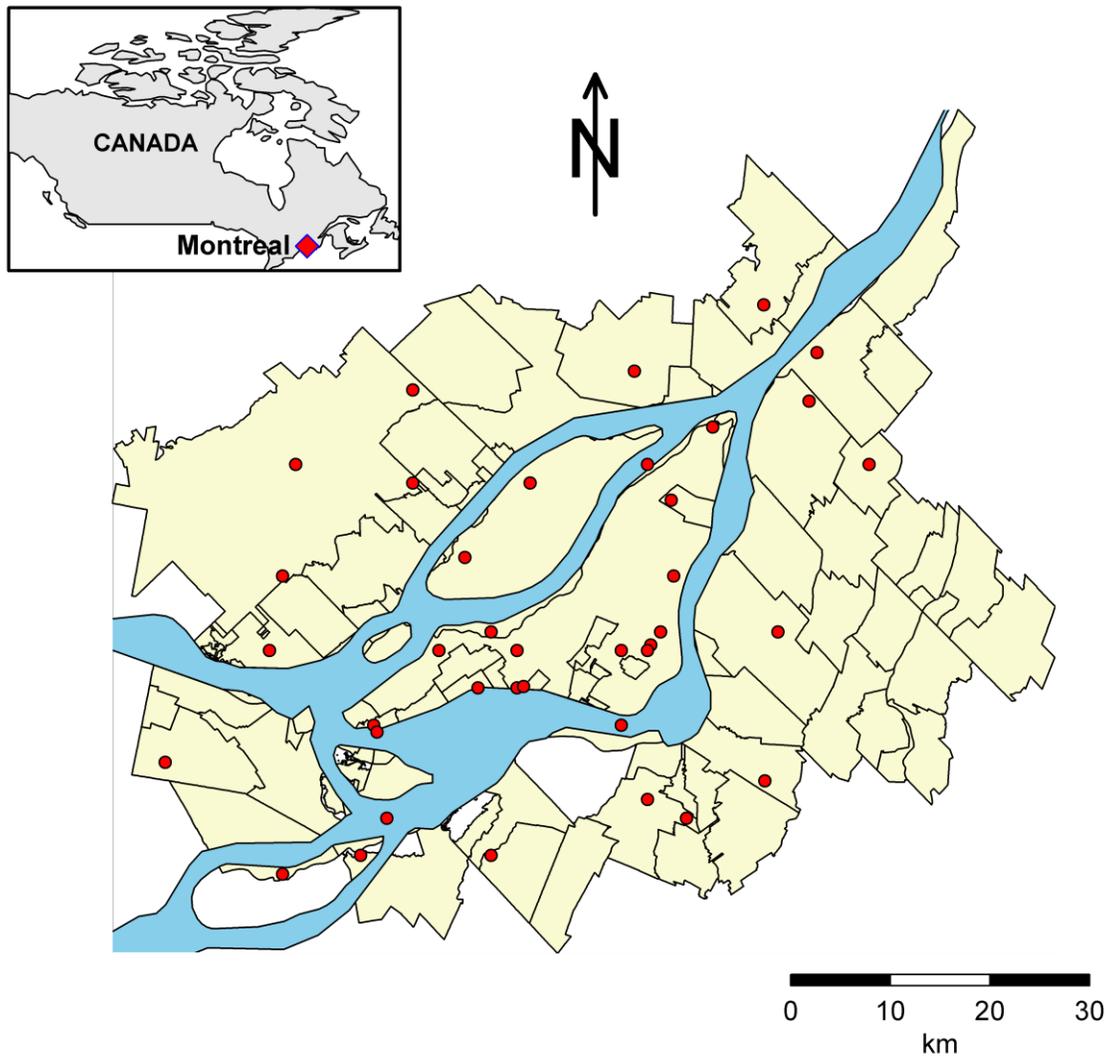

**Figure 2: Map showing the municipalities of the Montréal census metropolitan area along with the weather stations used for obtaining the temperature series (red dots).**

## 3.1. Data

The study region is the Montreal's CMA in the province of Quebec, Canada, shown in Figure 2. This area is the densest population basin of the province of Quebec, allowing for enough cases to be recorded in a small area where the weather can be considered homogeneous.

The health issue considered in this study is the daily mortality from cardiovascular diseases for the period 1981 to 2011 inclusively. The total population of the region was around 3 825 000 in the 2011 census. Cardiovascular diseases include ischaemic heart diseases (code I20-I25 in the tenth version of the international classification of diseases, ICD-10), heart failure (code I50), cerebrovascular disease, and transient cerebral ischaemic attacks (codes G45, H34.0, H34.1, I60, I61, I63 and I64). Corresponding codes ICD-9 were used for the period before 2000.

The environmental predictor is the daily mean temperature, which is the most studied variable in climate-related epidemiology (*e.g.* Gasparrini et al., 2015 and references therein). The temperature time series used is the spatial mean of the temperature measured at all the stations located in the Montreal's CMA. Note that kriging has also been considered to interpolate air temperature, but the results were similar to using the spatial mean in that region (Giroux et al., 2013).

### 3.2. Comparison to non-aggregated response

It is of interest to assess the practical differences between aggregating the response and using it directly. Since DLNMs are the most popular models in temperature-related health studies, the comparison is made through the use of DLNMs. Hence, three models are compared: i) classic DLNM estimated with non-aggregated $y_t$ response as a benchmark (model "C"), ii) DLNM estimated with an aggregated response using a 7-day moving average on future values ("MA") and iii) DLNM estimated with an aggregated response using a 7-day moving average on future values and taking into account the created temporal dependence (*i.e.* following Eq. (4), "MA-TS"). The 7-day moving average is used here as a starting example since it is the simplest and most commonly used aggregation scheme. Sensitivity of results to the aggregation scheme is checked in a subsequent section. The model MA is also applied to assess the benefits of taking the temporal dependence into account in a real-world case study.

The estimation of the three DLNMs has been designed similarly as in Gasparrini et al. (2015). Each model contains a smoothed time component estimated through natural cubic spline with 8 degrees of freedom per year, in order to control for unmeasured confounders. Note that confounding variables, such as humidity and air pollution, are not added in the considered models since the primary goal is to compare them. Because confounding variables are the same for all models, they should not change the comparison result. The DLNM surface is estimated by using a B-spline basis in both the temperature and the lag dimensions. The knots are placed at the $10^{th}$, $75^{th}$ and $90^{th}$ percentiles of the temperature variable and are equally placed on a log scale in the lag dimension, with maximum lag at 21 days. In addition, the classic model C contains an indicator variable for the day-of-week and is modelled as quasi-Poisson to account for over-dispersion. The other models (MA and MA-TS) are considered Gaussian as explained in section 2.2.

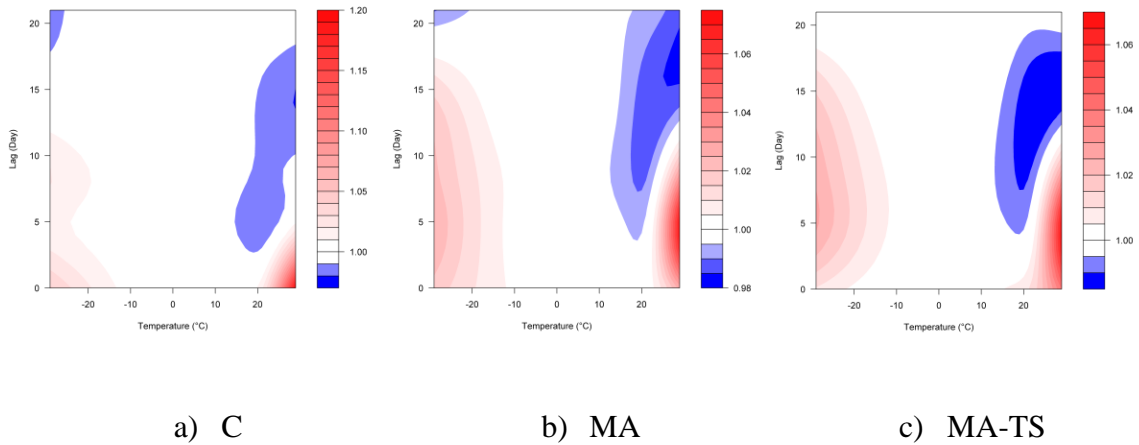

    a) C             b) MA             c) MA-TS

**Figure 3: Plot of relative risks along lag and temperature for models C (classical), MA (moving average) and MA-TS (moving average with time series regression).**

Figure 3 shows the surfaces reflecting the estimated effect of temperature on mortality through DLNM for the three models C, MA and MA-TS. Surfaces are presented as relative risks (RR), which are the exponentials of surfaces. The surface for C (Figure 3a) is consistent with the literature (*e.g.* Goldberg et al., 2011). It shows a comfort zone between -10°C and 20°C, an acute effect of hot temperature until three days lag and RRs under 1 for lags between 10 and 15 days, indicating a harvesting effect. In addition, a slight mortality excess of cold temperature (lower than -15°C) is found for lags between 0 and 10 days. The surface of the MA model (Figure 3b) is a smoothed version of the surface of the one of the C model. The effect of both heat and cold are longer and with less amplitude (RR not higher than 1.06). The surface obtained for the MA-TS model (Figure 3c) has the same shape as the previous one. It outlines more the latent effect of both cold and hot temperatures.

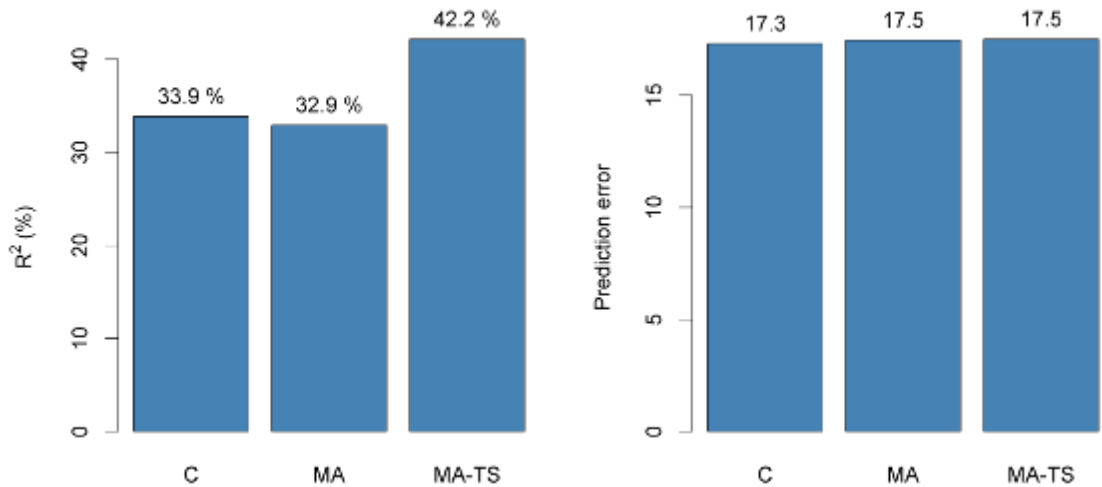

a) $R^2$     b) Cross-validated prediction error

**Figure 4: Numerical performance comparison between classical DLNM (C), DLNM with 7-day moving-average response (MA) and time-series regression with 7-day moving-average response (MA-TS).**

Figure 4 shows performance criteria $R^2$ and CV detailed in section 2.3. $R^2$ values (Figure 4a) are very similar between C and MA models (respectively 33.9 % and 32.9 % of explained variance), consistently with the similar surfaces found. However, the MA-TS model presents the highest $R^2$ value (42.2 % explained variance) showing the relevancy of the ARMA modelling of residuals. To compute CV values, Bergmeir and Benítez (2012) recommend to remove long-term trend and seasonality from the data before the analysis, instead of including a smooth time component as done in epidemiology. CV criteria (Figure 4b) show very similar values of prediction error when the response is aggregated (a value of 17.3 for model C versus values of 17.5 for both MA and MA-TS models). However, the difference is lower than the standard error of CV values which is around 0.5 meaning that a definitive increase of prediction error cannot be concluded.

### 3.3. Choice of the aggregation scheme

The previous section suggests that applying the model of Eq. (4) with a response aggregated through a 7-day moving average results in a better fit than a classical model, as well as providing other insights on the relationship especially at cold temperatures. The question is now: how do other aggregations behave and what is the better choice for $H$? Hence, the goal of this section is to compare the aggregation models listed in section 2.1 (and shown in Figure 1).

Figure 5 shows the performance criteria detailed in section 2.3 for each aggregation model with $H$ values between 3 and 21. Overall, the $R^2$ scores (Figure 5a) are higher for the lowest value of $H = 3$ and decrease as $H$ increases. The aggregation showing the highest $R^2$ values is the Epanechnikov kernel. However, its $R^2$ values are only 5 points higher than those of the Michels kernel on future values. The MA aggregations show the lowest $R^2$ values, showing the relevancy of considering non constant weights for aggregation.

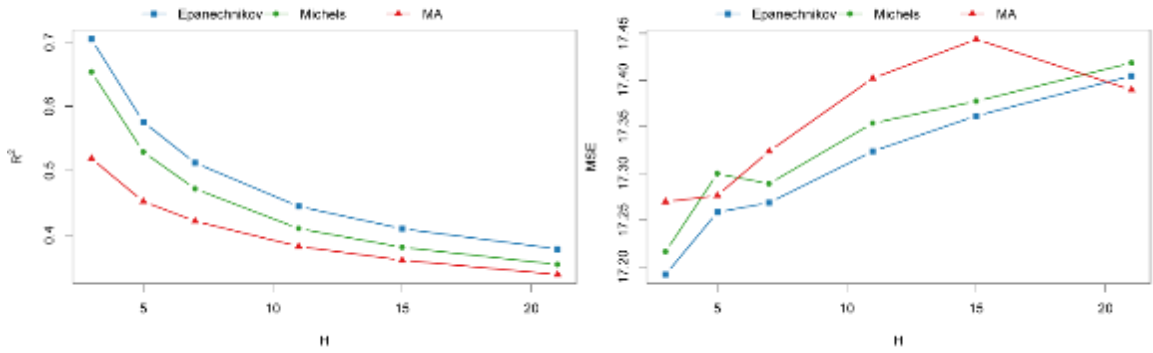

a) $R^2$         b) Cross-validation

**Figure 5: Performance criteria values for 3 different aggregations and different values of $H$ between 3 and 30.**

Similarly to the $R^2$ values, the CV values (Figure 5b) are the lowest (indicating best predictive performances) for $H = 3$ and slightly increase with $H$. The best aggregation according to the CV criterion is still the Epanechnikov kernel in opposition to the MA aggregation. The latter shows the highest prediction errors. However, the differences are still lower than the standard error of the CV values (around 0.5).

It has been stated that the Epanechnikov kernel presents the best performances of all the aggregation models considered in the present paper. Therefore, Figure 6 shows the DLNM surfaces obtained with Eq. (4) when the response is aggregated through the Epanechnikov kernel with $H = 3$ (Figure 6a) and $H = 7$ (Figure 6b). Note that many surfaces obtained with other aggregation schemes were very similar to those in Figure 6. Both surfaces are very similar to the surface of the model MA-TS (Figure 3c), by showing important RRs for the coldest temperatures around a 5-day lag. The main difference is that the RRs associated to the acute effect of extreme heat are increased, which could mean that the whole response to heat wave expositions is captured through the Epanechnikov kernel on future values.

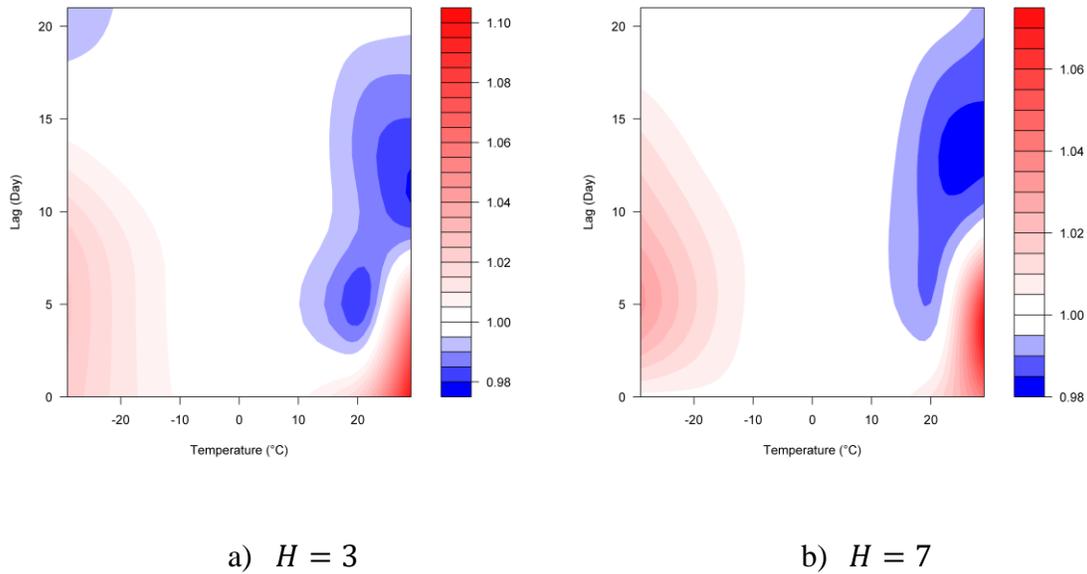

a) $H = 3$                                b) $H = 7$

**Figure 6: DLNM surfaces estimated when the response is aggregated through the Epanechnikov kernel.**

In conclusion, following the results of section 3.2, the present section shows that aggregating with the Epanechnikov kernel on future values (for which the shape is shown in Figure 1) is the most adapted aggregation scheme for temperature-related cardiovascular mortality studies in the context of Montréal, Canada. The Michels kernel on future values shows also good performances. Indeed, both improve the fit of the model and present a shape adapted to the mortality response of a temperature exposition.

## 4. Discussion

The CVD mortality and temperature data were used to compare DLNM without aggregated response to DLNM with aggregated response, with and without modelling the created temporal dependence. Results show that when the temporal dependence is not modelled, results are quite similar between aggregated response (model MA) and non-aggregated response (model C), although the former smooth the relationship. For the latter, it is important to note that the results and interpretation are very similar to what is found in the literature for Montréal such as in Doyon et al. (2008) as well as Goldberg et al. (2011).

The fit quality increases when the temporal dependence in residuals is modelled. In this case the surface slightly differs from the surface of model C. Indeed, the surface of the MA-TS model indicates a greater influence of cold temperature with a lag comprised between 5 and 10 days. This suggests that aggregating the response allows to obtain a signal at lower frequencies, which is not visible when considering the response directly. This also outlines the important effect of cold, as mentioned by Gasparrini et al. (2015). This is later confirmed by the surfaces obtained when the response is aggregated through the Epanechnikov kernel.

In their study, Sarmento et al. (2011) considered only 7-day centered moving averages. The present paper investigates other possible local aggregations with different window sizes. The aggregations investigated are MA, kernel smoothing with Epanechnikov and Michels kernels considering only future values (*i.e.* $i > 0$ in Eq. 1). The differences between aggregations are small which is logical since, in kernel smoothing, it is established that the choice of $H$ is much more important than the choice of the kernel (*e.g.* Wand and Jones, 1995). However, the Epanechnikov kernel shows slightly better performances than the others ones. As illustrated in Figure 1, this aggregations attributes maximum weight at the center of the window, and slightly decreases with higher lags, which represents well the physiological adaptation (*e.g.* Liu et al., 2015). The better performances for aggregation with small windows suggest that, in the case of cardiovascular mortality, aggregating the response is needed to remove some noise in the data, but a too large window results in the loss of information about acute effects, which are very important ones.

In this study, other aggregation schemes were considered, namely spectral smoothing with wavelets (Daubechies, 1992) and empirical mode decomposition (Huang et al., 1998). However, these aggregations do not have the property of being local (*i.e.* they use the entire series for aggregation), which violates the assumptions of hv-block CV and are less interpretable than the aggregation considered in the paper. In addition, they have shown poor results in comparison to the other aggregations. The results can nonetheless be found in the technical report (Masselot et al., 2016). Loess aggregation (Cleveland and Devlin, 1988) has also been considered during the study, but showed similar results to the Epanechnikov kernel and thus has not been presented here.

The proposed methodology is not intended to replace classical models but to be used in complement or in special cases. On the one hand, aggregating the response is helpful when a large amount of noise is suspected in the response and to study the relationship at a longer term than a daily basis. This is the case for areas with few mortality occurrences, such as northern areas in Canada, in which the temperature-related signal is dominated by the noise (Chebana et al., 2012). The aggregation could also be helpful in studies dealing with exposures with less acute effects such as humidity on cardiovascular mortality during spring (*e.g.* Masselot et al., 2018). On the other hand, there is a loss of information in the aggregation process, especially concerning health peaks. Dedicated extreme methods are important to study the extreme acute effects of an environmental exposure on a health issue (*e.g.* Chiu et al., 2016).

Although the proposed modelling allows considering nonlinear models such as the DLNM, the modelling is restricted to models that can be expressed as a linear combination of basis functions. Other estimation methods such as penalized models (*e.g.* Ridge and Lasso regressions), which are very useful for several correlated predictors, would necessitate the development of dedicated models.

## 5. Conclusions

The present paper proposes to aggregate the health response in environmental epidemiology studies, in order to reduce the importance of noise in the health data. The proposed methodology consists in aggregating the response and then applying a time series regression model to account for the temporal dependence created by the aggregation. This model is general and therefore not limited to linear regression and allows the use of DLNMs. The proposed methodology is then applied to the practical issue of temperature-related cardiovascular mortality.

Results show that aggregating the response and modelling the temporal dependence leads to an increase in the fit quality. In addition, it outlines the longer term nature of the relationship during winter in Montreal, which is not as obvious using the classical DLNM. It is shown and argued that using an asymmetric aggregation (the Michels kernel or the Epanechnikov kernel on future values) with a small window leads to better results.

The obtained results and findings in the present paper are valid only for the case of temperature-related cardiovascular mortality in Montréal. Hence, these results cannot be generalized to other locations or variables, as illustrated by the diversity of results in Gasparrini et al. (2015). Therefore, as a future perspective, it could be of interest to apply the proposed methodology to other health issues and other locations.

## Acknowledgements

The authors are thankful to the *Fonds Vert du Québec* for funding this study and to the *Institut national de santé publique du Québec* for data access. The authors also wish to thank Jean-Xavier Giroux (INRS-ETE) for his help with database building, Yohann Chiu (INRS-ETE) for all his relevant comments during the project as well as two anonymous reviewers for their helpful comments in improving the quality of the paper. All the analyses were performed using the *R* software (R Core Team, 2015) with the package *forecast* (Hyndman, 2015). The R codes are freely available upon request.